\documentclass{mn2e}
\title[Gravitational lenses in the COSMOS fields]
{Gravitational lenses and lens candidates identified from the COSMOS field}
\author[N. Jackson]
{Neal Jackson
\\
Jodrell Bank Centre for Astrophysics, University of Manchester,
Turing Building, Oxford Road, Manchester M13 9PL\\}
\input psfig.tex
\begin{document}
\maketitle
\begin{abstract}
A complete manual search has been carried out of the list of 285423
objects, nearly all of them galaxies, identified in the COSMOS field 
that are brighter than $I=25$. Two certain and one highly probable 
new gravitational lenses are found, in addition to the lenses and 
candidate lens systems previously found by Faure et al. (2008). A
further list of 112 candidate lens systems is presented. Few of these
are likely to be true gravitational lens systems, most being
star-forming rings or pairs of companion galaxies. It is possible
to examine of order 10$^6$ objects by eye in a reasonable time, although
reliable detection of lenses by such methods is likely to be possible only with
high-resolution data. The loss of completeness involved in a rapid
search is estimated as up to a factor of 2, depending on the morphology
of the lens candidate.
\end{abstract}


\large

\section{Introduction}

There are a number of reasons why gravitational lenses, systems in which
a distant quasar or galaxy are multiply imaged by foreground galaxies, 
are important.
Because lensing is a purely gravitational effect, the inverse problem
can be solved in order to reconstruct the lensing mass distribution from
the observed images, independently of the nature of the matter in the
lens. Variable lensed sources can be monitored for time delays between
variations of the images, leading to determination of the Hubble constant
(Refsdal 1964, for reviews see e.g. Courbin 2003, Kochanek \& Schechter
2004, Jackson 2007); non-correlated image variability allows us to study 
microlensing by stars in relatively high-redshift galaxies (e.g. Agol \&
Krolik 1999, Wambsganss 2001, Ofek \& Maoz 2003, Morgan et al. 2008); lensing
statistics, along with a cosmological model, allow constraints on
galaxy evolution (e.g. Chae \& Mao 2003, Ofek, Rix \& Maoz 2003, 
Matsumoto \& Futamase 2008); and the magnification induced by lenses allows us
to study intrinsically fainter objects.

Since the discovery of the first gravitational lens by Walsh et al. (1979),
over 100 lens systems have been discovered. The CASTLES compilation website
(Kochanek et al. 2008) lists 100 cases in which quasars are multiply
imaged by foreground galaxies. In addition, large numbers of lens systems
in which the lensed sources are extended galaxies are now being produced
by searches such as the SLACS survey (Bolton et al. 2006). Searches for
lensed arcs in large-area sky surveys are now beginning to be
successful; for example, Cabanac et al. 2007 report progress in the use
of the Canada-France-Hawaii Telescope Legacy Survey for this purpose.
Recently, Faure et al. (2008) report the use of the Cosmic Evolution
Survey (COSMOS, Scoville et al. 2007) 
as a means of identifying gravitational lenses. COSMOS 
is a 2 square degree area in which HST/ACS imaging, together with space-
and ground-based followup by telescopes at many different wavelengths, 
has identified over two million objects.

In the COSMOS survey there are 285423 sources brighter than 25th
magnitude in the ACS $I$-band observations. Faure et al. (2008)
considered a subset of 9452 of these objects. This sample was chosen to 
be the most likely to contain gravitational lens systems. The most likely
lens systems are those at moderate redshift ($0.2<z<1.0$), with 
intrinsically high luminosity ($M_V<-20$) and those spectrally 
classified as early type galaxies. They find 20 good candidates, 
many of which are likely on morphological grounds to be gravitational 
lenses. In addition, they list 47 other objects in which detection of a 
single arc gives some indication of lensing by the primary galaxy.

In the present work, all of the COSMOS catalogue images are manually inspected.
Two definite gravitational lenses are found, together with a third
highly probable system. The search strategy is described in section 2,
and the definite lens systems are presented in section 3. Finally in
section 4, implications and future prospects are briefly
discussed.

\section{Examination of the COSMOS images}

\subsection{Method}

Pseudo-colour images, 5\farcs25 on a side, were made in a similar way to
those of Faure et al. 2008, using the COSMOS catalogue (Capak et al. 2007)
and the ACS images from the COSMOS database for intensity in each pixel,
and Subaru images in $B$, $r^+$ and $i^+$ (Taniguchi et al. 2007) for the
colour coding. Colour gradients and lookup tables were optimized for fast
visual inspection. A displayed intensity level $I$ related to counts $S$
by $I\propto S^{0.9}$ was found to give the best results. The maximum 
intensity level was set to the counts in the eighth brightest pixel in 
the central 0\farcs5$\times$0\farcs5, or 0.2 counts$\,$s$^{-1}$, whichever
was the smaller. This condition burns out the central region of about 5\%
of the brighter galaxies, but a large majority of these will appear in the
compilation of Faure et al. (2008), or be too close to have a
significant lensing cross section. The colour indices for each pixel
were calculated by first normalising each pixel in each of the three
colour images to the average pixel value for the colour image, and
taking the output value for each pixel as the normalised value for that
pixel raised to some power $\beta$. If we denote the three output colour
values for a pixel by $b$, $g$ and $r$, the intensity for the level of
the output blue image was $3ab/(r+g+b)$ where $a$ is the count level 
from the high-resolution ACS image, $3ag/(r+g+b)$ for the green level
and $3ar/(r+g+b)$ for the red image. A higher value of $\beta$ gives
more colour contrast, and in practice $\beta=2$ was found to be a useful
value.

Cutouts were made around the position of each object in the catalogue,
and the images were then mosaiced into 6$\times$4 frames and examined by
eye using the ImageMagick\footnote{ImageMagick is available under the
GNU Public Licence from http://www.ImageMagick.org.} software. Between
8000 and 10000 objects per hour can be examined in this way.

Criteria for
regarding galaxies as candidate lenses included multiple images,
particularly those corresponding to plausible lensing configurations,
and structures similar to lensed arcs. In the case of multiple-image
systems, the easiest systems to recognise are the four-image (``quad'')
systems produced by sources within the astroid caustic of an elliptical
lens. The only reason for failure to recognise such systems should be
faintness ($I_{814}>25$) of one or more of the images. Two-image systems,
especially those with a faint secondary, are much more difficult. An
attempt was made to include such systems, but two-image
systems with pointlike sources are likely to be missed unless both
components are well above the $I_{814}=25$ level. For typical flux
ratios of about 5, the survey will therefore be insensitive to double
systems fainter than $I_{814}=22-23$ unless they are accompanied by arc
structures, and for this reason it is not surprising that all of the 
likely new lenses are quad systems. Because of the good
resolution of the ACS images, components of lenses can be detected to
within 200~mas of the lensing galaxy, and due to the 5\farcs25 size of the
cutouts, lenses of Einstein radii up to 2\farcs5 can be detected. This
range corresponds to nearly all lenses produced by single galaxies
without substantial assistance from a cluster (e.g. Browne et al. 2003).

Lensed arcs from extended background objects are likely to form the
majority of lensed systems. In principle these can be detected easily,
the main criteria being tangential extension with respect to the lensing
galaxy and curvature of the arc. The main confusion in this case is with
nearby galaxies, interacting with the primary object and stretched by
tidal effects associated with the interaction. Objects which have more
credibility as lens candidates are those with relatively thin, long arcs,
particularly those with significant colour differences from the primary
galaxy and without a significant colour gradient across them. In
practice, selected candidates have typical length-to-width ratios of
about 3:1 or greater. In lens systems where the arcs are extended into an
Einstein ring, such systems are most likely to be rejected due to 
confusion with star-forming rings in the primary galaxy. Colour information
is sometimes of limited help here, as both lensed background objects and
star-forming rings are expected to be bluer than the predominant light 
from the primary. The compromise between including false positives and 
rejecting lenses is at its most subjective in these cases.

From the list of lens candidates, objects already identified as possible
lens systems by Faure et al. (2008) are excluded. The remainder are
divided into two categories: candidates (possible or probable lenses)
and likely lenses (very likely or certain lenses). In all, 112 candidates
and 3 likely lenses survive this selection process. The lenses are
discussed further in Section 3, and the candidates are presented in Fig. 1,
with the coordinates of each object given in the figure itself. The
candidates vary considerably in credibility. For example, 095806+021726
is a weak candidate due to the shortness of the arc; the extended
component could plausibly be a companion galaxy. The arc in
100000+021545, although relatively long, is also a weak candidate because
of its lack of curvature. On the other hand, arcs such as those in
095950+022057 are morphologically much stronger lensing candidates,
although the possibility remains that they could also be star-forming
rings within the main galaxy. Possibly the strongest candidate,
100141+021424, has an Einstein ring-like structure, but there is a
colour gradient across the ring. This does not necessarily rule the
object out as a lens, because it may be a lensed ring superimposed upon
colour gradients within the lensing galaxy.

\begin{figure*}
\psfig{figure=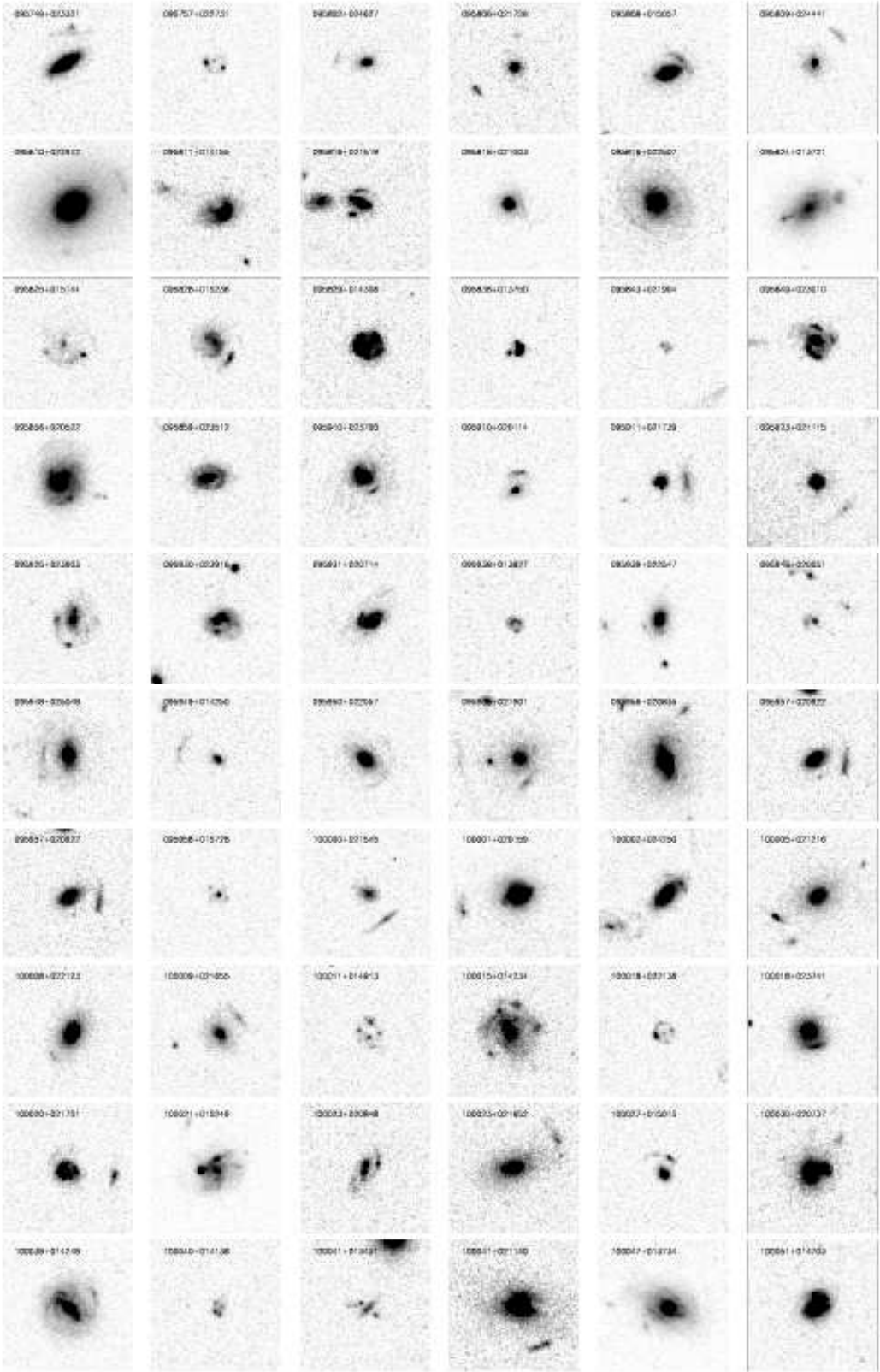,width=14cm}
\caption{Candidate lens systems from the manual examination of COSMOS
images, excluding candidates already presented by Faure et al. (2008). 
Most are single arc systems which may also be star-forming
rings in the galaxy.}
\end{figure*}

\setcounter{figure}{0}
\begin{figure*}
\psfig{figure=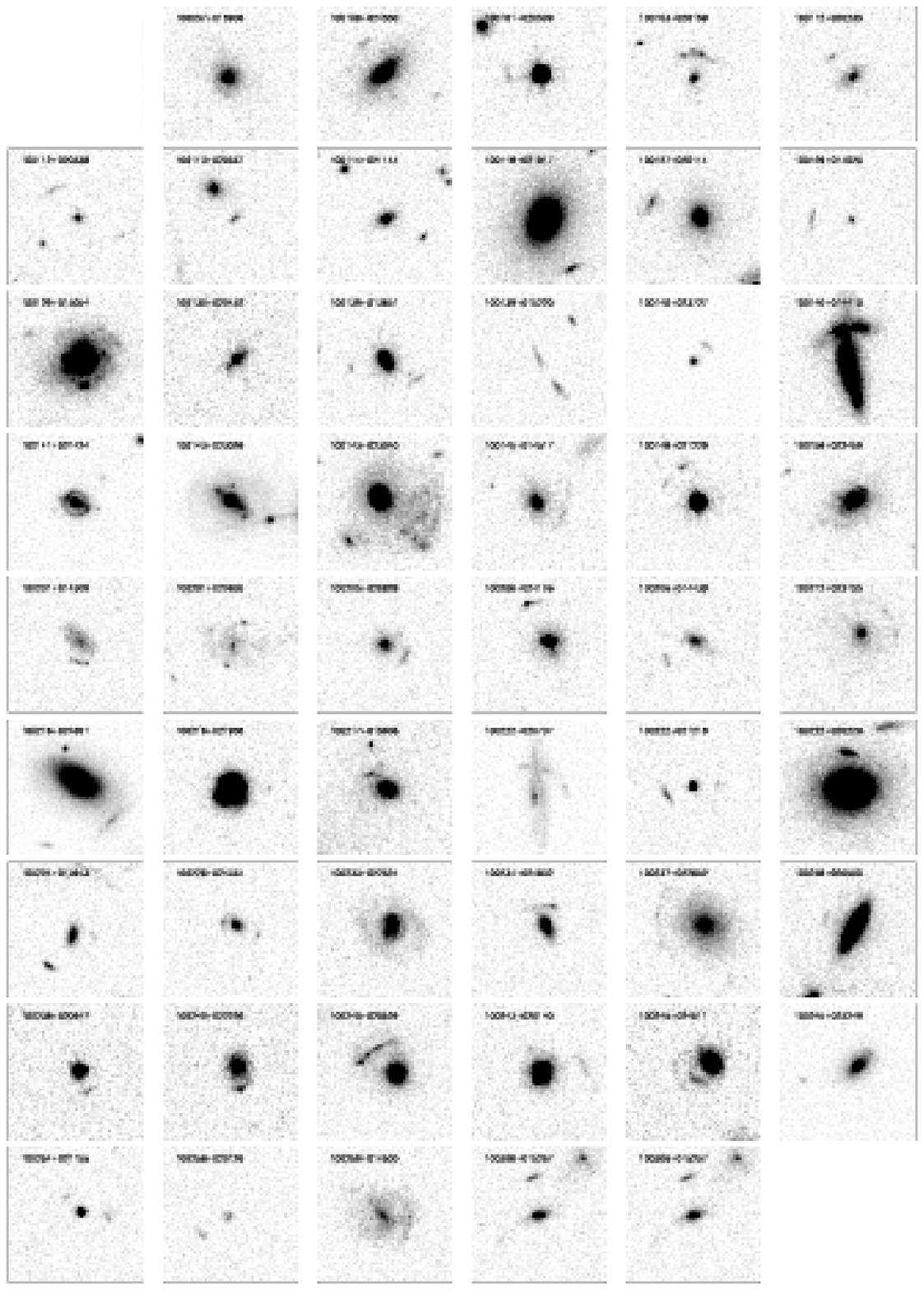,width=16.5cm}
\caption{Candidate lens systems from the manual examination of COSMOS
images (continued).}
\end{figure*}

\subsection{Completeness of the examination}

Although this examination of the COSMOS images is comprehensive in the
sense that every object has been looked at, it has been optimized for
speed rather than completeness. One obvious incompleteness is that
bright galaxies are examined more cursorily than in the work of Faure et
al. (2008), and in particular no attempt has been made to subtract
elliptical isophotes from these objects. 
Here we assess the completeness (extent to which real gravitational lenses
are present in the sample of candidates) by comparison with Faure et al.
(2008). Because Faure et al. present a much more careful manual
examination of a much smaller number of candidates, their results are
likely to be closer to a complete lens candidate sample; the search presented
here was deliberately performed without reference to the candidate lists
of Faure et al.

Faure et al. present a list of 20 ``best systems''. Of these, one
(095737+023424) is missing from our sample because it does not have an
$I_{814}$ magnitude in the COSMOS catalogue, being outside the region of
ACS coverage. Of the remaining 19, 10 are detected by the manual search 
presented here. ACS images of the 10 systems found and the 9 not found 
are presented in Figure 2.

\begin{figure*}
\psfig{figure=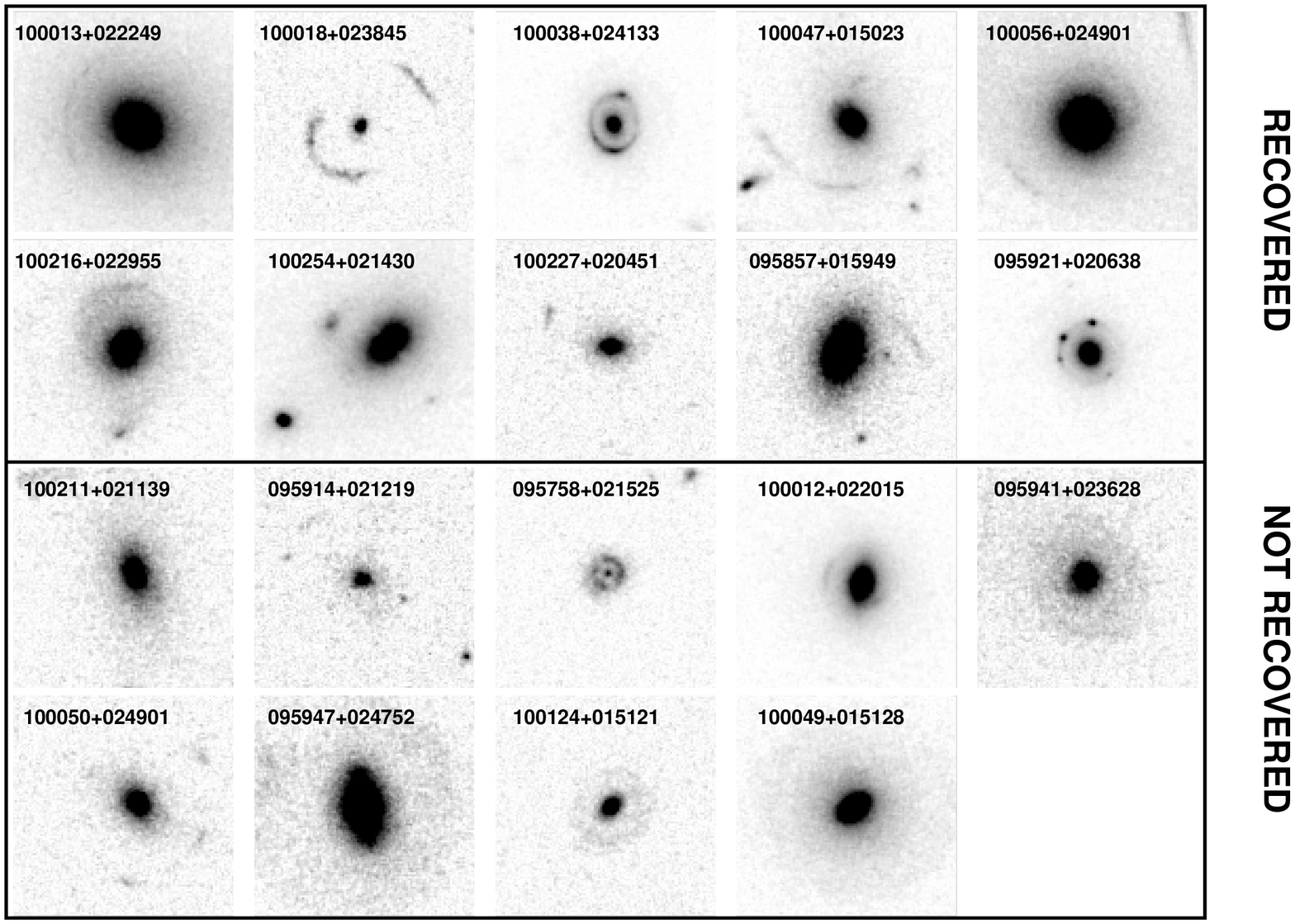,width=15cm}
\caption{The 19 ``best systems'' of Faure et al. which have $I_{814}$
magnitudes in the COSMOS field. 10 of these 19 are recovered in this
survey (top two rows), including the majority of the ``obvious'' lenses.
The remainder are not recovered for a variety of reasons, including
lensed arcs further than 2\farcs5 from the primary and thus outside the
cutout window, relatively faint
rings or extended structure, or morphology judged during the independent
search to be due to non-lensing structures; see text for further details.}
\end{figure*}

In one of the nine cases not recovered as a candidate, the ring 
structure of the candidate was missed
due to its large diameter and the fact that, for ease of rapid
examination, the 5\farcs25 cutout size is smaller than the 10\arcsec\  
used by Faure et al. In the other eight cases, there are two main reasons
for non-detection of the lens. The first is faintness of some of the arc
structures, which can be missed during rapid inspection (e.g.
100049+015128). The second reason is the appearance of structures
which, while they are consistent with lensing, may also be consistent
with structures such as those seen in polar ring galaxies or galaxies
with ring-shaped regions of star-formation (e.g. 095941+023628).
The inclusion of all such objects from the whole catalogue would have
resulted in a considerably increased number of candidates, and such
objects are accordingly discriminated against compared to multiple-image
lenses.

Encouragingly, nearly all of the Faure et al. candidates which are 
``obvious'' candidates -- those which clearly contain subsidiary flux 
concentrations well above the noise, and which cannot reasonably be 
explained by any mechanism other than lensing (e.g. 095921+020638) 
are recovered by this survey. Quantification of completeness is subjective,
and depends on individual judgement about the candidates that were missed.
For what it is worth, it is the author's opinion that at least three of the 
candidates of Faure et al. that were missed should definitely have been 
in the sample, corresponding to a completeness level of $\leq$10/13, and
that two of these three were missed due to the difficulty 
of adjusting the colour scheme and lookup table to clearly distinguish 
relatively low signal-to-noise features, while still maintaining a rapid
rate of inspection. This is an obvious area to optimize further.

We can also investigate the extent to which candidates are recovered by
a second examination of a small subsample. Again, the recovery fraction
depends on the ``quality'' of the lens candidate. For ordinary
candidates (e.g. single arcs without long tangential extension, or weak
multiple images) the recovery rate is about 50\%, but this increases
with increasing candidate quality; for example, in Fig. 1 the systems
100141+021424 (clear ring) and 100205+020808 (extended arc) are recovered,
but 100148+021229 (small arc-like feature) and 100114+021144 (possible
arc-like feature) are not. It is unlikely that
further ``obvious'' lens systems such as J100410.12+020040.9 or 
J095930.94+023427.7 remain within the dataset. The lack of further
obvious candidates vindicates Faure et al.'s assumption that their
selection of intrinsically luminous objects at moderate redshift is an
efficient way of finding lenses.

\section{New gravitational lenses}


Two new objects are found which are clearly gravitational lens systems,
based only on morphological and colour evidence. A third object has a
morphology which resembles an Einstein Cross configuration, similar to the lens
J2237+0305 (Huchra et al. 1985). Although this object is likely to be a gravitational
lens, it requires confirmation. Each object is discussed separately.

\begin{figure*}
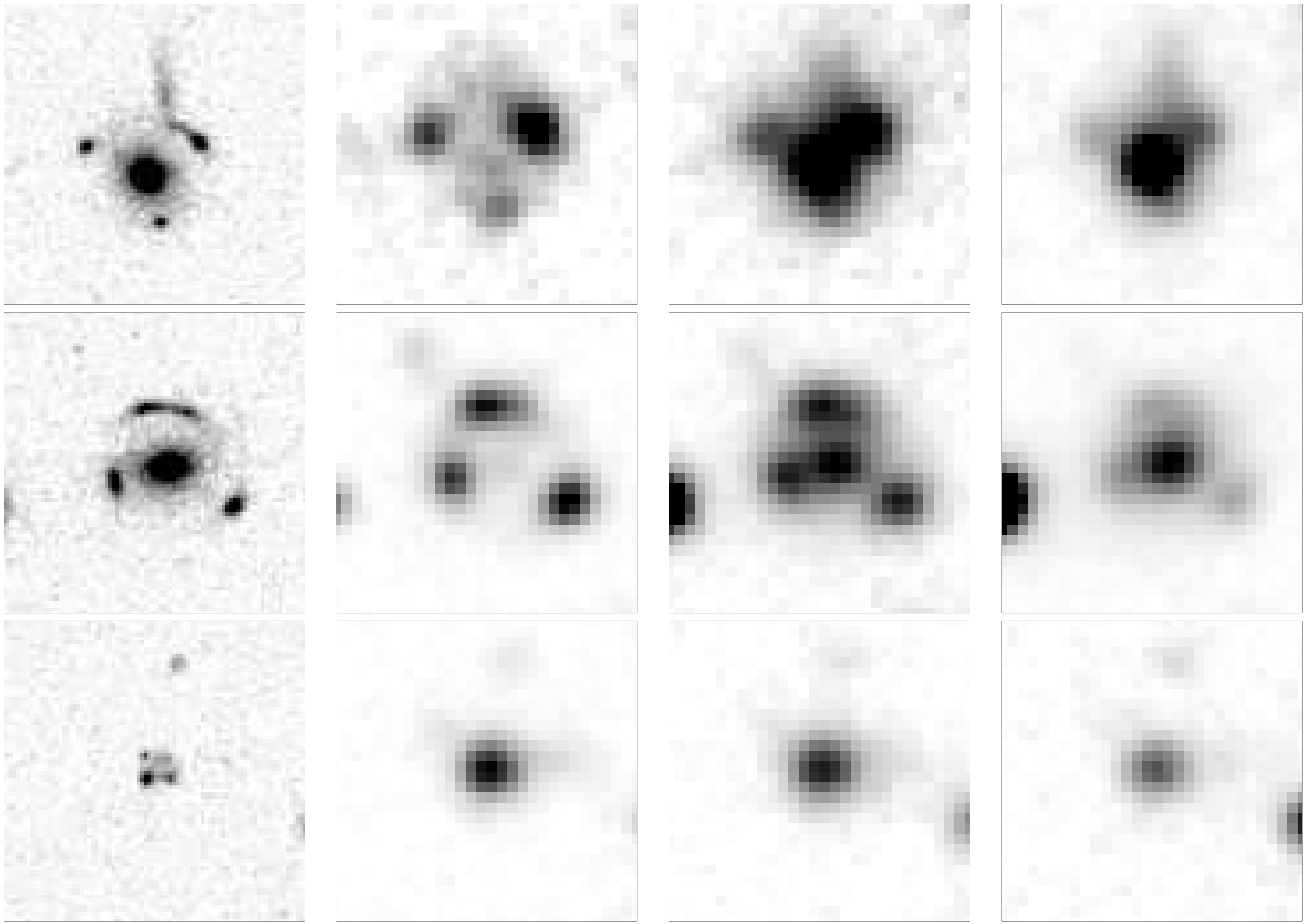

\begin{tabular}{cccc}
\psfig{figure=1185017_acs.fits.ps,width=4cm,angle=-90}&
\psfig{figure=1185017_blue.fits.ps,width=4cm,angle=-90}&
\psfig{figure=1185017_green.fits.ps,width=4cm,angle=-90}&
\psfig{figure=1185017_red.fits.ps,width=4cm,angle=-90}\\
\psfig{figure=2403466_acs.fits.ps,width=4cm,angle=-90}&
\psfig{figure=2403466_blue.fits.ps,width=4cm,angle=-90}&
\psfig{figure=2403466_green.fits.ps,width=4cm,angle=-90}&
\psfig{figure=2403466_red.fits.ps,width=4cm,angle=-90}\\
\psfig{figure=135518_acs.fits.ps,width=4cm,angle=-90}&
\psfig{figure=135518_blue.fits.ps,width=4cm,angle=-90}&
\psfig{figure=135518_green.fits.ps,width=4cm,angle=-90}&
\psfig{figure=135518_red.fits.ps,width=4cm,angle=-90}\\
\end{tabular}
\caption{HST and Subaru images of the new lens systems
J100140.12+020040.9 (top), J095930.93+023427.7 (middle) and
J100126.02+013714.5 (bottom). The columns from left to right are HST/ACS
(I-band), Subaru $B$, Subaru $r^+$, and Subaru $i^+$. Images are
4\farcs3 on a side; North is at the top and East to the right.}
\end{figure*}

\subsection{COSMOS~J100140.12+020040.9}  

Object J100140.12+020040.9 (Fig. 3) is a galaxy with 
an $I_{814}$ magnitude of 21.86. 
It is a clear example of a four-image gravitational lens system. A pair of
merging images is located 0\farcs85 NW of the centre of the lensing
galaxies, and two other images are located approximately 0\farcs9 NE and
0\farcs7 slightly W of S. All of these images have similar colours, and
are much bluer than the lens galaxy. Of the two merging images, the 
southeastern one is faint, possibly due to reddening by the second 
galaxy or alternatively to microlensing. The light is dominated by the
galaxy in the red Subaru $i^+$ image, and by the lensed images in the
blue ($B$) image. The isophotes of the lens galaxy appear almost
circular. 

A second galaxy is visible about 1\farcs2 north and slightly west of the
main galaxy; it is highly elongated in an N--S direction.
There are a number of nearby galaxies in what appears to be a
small group, the nearest being about 4\arcsec\  away to the NW.

The light profile of the system was modelled using the GALFIT software
of Peng et al. (2002), fitting Sersic profiles to both the primary and
secondary galaxies, and point spread functions generated with TinyTim
(Krist 1993) for the images. The primary is modelled as an $I=21.8$ object of
Sersic index 3.1 (where 1 represents an exponential disk and 4 a de
Vaucouleurs profile) and is likely to be a standard elliptical galaxy.
The photometric redshift of 0.81 in the COSMOS database (Mobasher et al.
2007) implies an approximately $L_*$ galaxy, but it is possible that
the photometric redshift may be affected by the combination of red
galaxy and blue images, and therefore that the galaxy redshift may be
significantly less than 0.81.

A model can be made of the system using a singular isothermal sphere for
the primary galaxy and an SIE for the second galaxy. The positions of
the four images, measured using the AIPS task MAXFIT to fit to the
centre of the light distributions, are given in Table 1. A fit to the four
observed images can be made by varying the Einstein radii of both
galaxies, the ellipticity of the second galaxy, plus external shear
(magnitude and position) in the system. The resulting model has no
degrees of freedom, although deeper observations of the arc system may
provide further constraints. Reduction of $\chi^2$ to 1 requires a small movement
(about 0\farcs02) of the main galaxy from the measured position; this is
achieved by allowing the position in both RA and Declination to move,
subject to a Gaussian penalty function with $\sigma$=0\farcs02.

\begin{table}
\begin{tabular}{ccc}
Offset in RA (arcsec) & Offset in Dec (arcsec) & Relative flux \\ \hline
$-$0.7255$\pm$0.001&+0.471$\pm$ 0.001 &1.00\\
+0.858$\pm$0.002& +0.430$\pm$0.002 &0.51$\pm$0.03\\
$-$0.189$\pm$0.003& $-$0.657$\pm$0.003 &0.42$\pm$0.04\\
$-$0.444$\pm$0.006& +0.692$\pm$0.006 &0.06$\pm$0.06\\ \hline
\end{tabular}
\caption{Positions and fluxes of images in COSMOS~J100140.12+020040.9.
Offsets are given from the measured position of the main lens galaxy.}
\end{table}

\begin{table}
\begin{tabular}{ccc}
Quantity & Main galaxy & G2 \\ \hline
Offset in x (\arcsec )&0.020&0.260\\
Offset in y (\arcsec )&0.005&1.109\\
Einstein radius (\arcsec )&0.787&0.068\\
Ellipticity &0 (fixed)&0.812\\
Position angle (deg)&--&7.32\\
External shear magnitude &0.069&\\
Shear angle (deg)&$-$11.79&\\ \hline
\end{tabular}
\caption{Parameters of the lens model for COSMOS~J100140.12+020040.9.
Offsets are given from the measured position of the main lens galaxy.}
\end{table}

\subsection{COSMOS~J095930.94+023427.7} 

COSMOS~J095930.94+023427.7 is a $I_{814}=21.76$ object with a COSMOS
photometric redshift of 1.21, and is clearly a four-image gravitational lens
system. Two merging images lie about 0\farcs8 of a lensing galaxy, and
two further images are present to the southeast and southwest, the
southwestern image being noticeably further from the lensing galaxy.
Again, the Subaru colour imaging clearly shows that the lensed images
have a similar blue colour, and the lensing galaxy dominates the light
in the $i^+$--band.

A lens model is more difficult to construct for this object. Using the
observed point-source positions (again estimated using MAXFIT) gives
eight constraints, and we can make a model with seven free parameters
assuming an isothermal profile for the lensing galaxy (Einstein radius,
ellipticity and position angle, external shear magnitude and direction,
and source position) which does not give a good fit: $\chi^2\sim 500$,
which is unacceptable even given the likely slightly optimistic errors.
Allowing the galaxy position to vary, again by a $\sigma=0\farcs02$
Gaussian penalty function, yields a better fit, at the cost of a large
offset to the North of the lensing galaxy and an unrealistically large
external shear ($\gamma=0.254$ is a shear which would be produced by an
isothermal perturber of the same luminosity as the primary and about
1\farcs5 away, or by a cluster of $\sim 1000$~km$\,$s$^{-1}$ within about 
an arcminute). This shear points in the direction of a bright galaxy in
the field, which is however 14\arcsec\ away from 
COSMOS~J095930.94+023427.7. 
The problems in fitting this source are reminiscent of
the difficulties in the four-image lens CLASS~B0128+437 (Biggs et al.
2004) where again a large shear is required, much larger than can
reasonably be ascribed to objects in the field.

\begin{table}
\begin{tabular}{ccc}
Offset in RA (arcsec) & Offset in Dec (arcsec) & Relative flux \\ \hline
+0.752$\pm$0.0050 & $-$0.1640$\pm$0.005 &1.00\\
$-$0.925$\pm$0.0050 & $-$0.5520$\pm$0.005 &0.93$\pm$0.03\\
$-$0.266$\pm$0.0050 & +0.7895$\pm$0.005 &0.63$\pm$0.03\\
+0.316$\pm$0.0050 & +0.8495$\pm$0.005 &0.76$\pm$0.03\\ \hline
\end{tabular}
\caption{Positions and fluxes of images in COSMOS~J095930.94+023427.7.
Offsets are given from the measured position of the main lens galaxy.}
\end{table}

\begin{table}
\begin{tabular}{cc}
Quantity & Main galaxy  \\ \hline
Offset in x (\arcsec )&0.004\\
Offset in y (\arcsec )&0.063\\
Einstein radius (\arcsec )&0.888\\
Ellipticity &0.516\\
Position angle (deg)&85.12\\
External shear magnitude &0.254\\
Shear angle (deg)&$-$17.42\\ \hline
\end{tabular}
\caption{Parameters of the lens model for COSMOS~J095930.94+023427.7.
Offsets are given from the measured position of the main lens galaxy.}
\end{table}

\subsection{COSMOS~J100126.02+013714.5}

COSMOS~J100126.02+013714.5 (Fig. 3) is an interesting source, but is
less obvious as a gravitational lens candidate. It is much fainter 
($I_{814}=23.73$) than the two definite lens systems. There are four bright
condensations, arranged in a cross and strongly reminiscent of Einstein
Cross lens systems such as Q2237+0305 (Huchra et al. 1985). Between
these there is clearly more extended emission. This is compatible with a
lensing galaxy which produces four lensed images, and also with a galaxy
which happens to have four star-forming regions in a configuration
resembling a lens system.

\section{Discussion and conclusions}

Th number of lens systems that we should expect in the COSMOS
survey is likely to be higher than the number of secure lens systems as
identified by eye. For example, Miralda-Escude \& Lehar (1992), in their
study of lensed arcs in optical surveys, predict an approximate number
of 100 per square degree in a survey to a depth of $B=26$, which for
blue objects roughly corresponds to the survey depth of the search 
undertaken in this work.
Since the COSMOS footprint is about 1.6 square degrees, this suggests
that over 100 lens systems should be present. In addition, Bolton et al.
(2008) report a total of $\sim$70 clearly identified lenses from the 
SLACS survey, which results from the study of $\sim$50000 spectra of 
luminous red galaxies (LRGs). Such galaxies are more likely than the 
average to be gravitational lenses, because of their larger lensing 
cross-section; nevertheless, even if the lensing cross-section of an 
average galaxy is a factor of 10 lower than an LRG we would expect 
several dozen lens systems. Most of the systems are therefore likely 
to be concealed within the single-arc systems of Faure et al. (2008) 
or those of Figure 1.

Direct confirmation of such lenses is not easy, because it requires
spectroscopy of $I=25$ arcs to determine redshifts. This is a
non-trivial task with an 8-m class telescope, and a programme is under
way to investigate the best candidates. In practice, however, it is
likely to be impossible to be substantially complete for lenses in a 
large, blind optical survey such as COSMOS.

In the future, it is likely that arc detection algorithms will be
applied to large-area surveys such as COSMOS and CFHTLS, and indeed
steps in this direction are described by Cabanac et al. (2007) and by
Seidel \& Bartelmann (2008). The latest algorithm to be used is an
automated robot which explicitly fits each image as a potential lens
system and adjusts the model to maximise source plane flux (Marshall et
al. 2008). Other methods include searches for multiple
blue objects around likely lens galaxies (Belokurov et al. 2007). 

Provided that high-resolution images are available, examination by 
eye of large samples is likely to be a valuable adjunct, both to 
provide example sets for arc detection algorithms, or even for neural 
network versions of these algorithms. The examination performed
in this work is crude, but it is surprisingly quick to perform; in fact, a 
sample of a million sources could be processed by one investigator with
about a month of dedicated effort.
The disadvantage is that many forthcoming surveys
will not have diffraction-limited resolution, making galaxy-scale lens
systems much more difficult to detect. 
Despite this, manual inspection of galaxies 
from forthcoming medium-deep surveys, such
as the VST and VISTA KIDS/VIKING surveys, each of which will image
several hundred thousand galaxies, will be useful. Lensing by
galaxies assisted by groups or clusters will be detectable; such
inspections will be more incomplete for lens systems such as those
presented here.

\section*{Acknowledgements}

The research was supported in part by the European Community's
Sixth framework Marie Curie Research Training Network Programme,
contract no. MRTN-CT-2004-505183 ``Angles''. I thank Ian Browne for
discussions on the paper and opinions on the candidate
lenses, and a referee for useful comments.

\section*{References}

\noindent Agol E., Krolik J., 1999, ApJ 524, 49

\noindent Belokurov V., et al., 2007, ApJ 671, L9

\noindent Biggs A.D., Browne I.W.A., Jackson N.J., York T., Norbury M.A., McKean J.P., Phillips P.M. 2004,  MNRAS, 350, 949. 

\noindent Bolton A.S., Burles S., Koopmans L.V.E., Treu T., Moustakas L.A. 2006,  ApJ, 638, 703. 

\noindent Bolton A.S., Burles S., Koopmans L.V.E., Treu T., Gavazzi R., 
Moustakas L.A., Wayth R., Schlegel D.J., 2008, astro-ph/0805.1931

\noindent Browne I.W.A., et al., 2003, MNRAS 341, 13

\noindent Cabanac R.A., Alard C., Dantel-Fort M., Fort B., Gavazzi R., Gomez P., Kneib J.P.,
LeF\'evre O., Mellier Y., Pello R.,et al. 2007,  A\&A, 461, 813. 

\noindent Capac P. et al., 2007, ApJS 172, 99

\noindent Chae K.-H., Mao S., ApJ 599, L61

\noindent Courbin F., 2003, astro-ph/0304497

\noindent Faure C., et al., 2008. astro-ph/0802.2174.

\noindent Huchra J., Gorenstein M., Kent S., Shapiro I., Smith G., Horine E., Perley R. 1985,  AJ, 90, 691. 

\noindent Jackson N., 2007, LRR, 10, 4

\noindent Kochanek C.S., Falco E.E., Impey C., Lehar J., McLeod B., Rix
H-W., 2008. http://cfa-www.harvard.edu/glensdata/

\noindent Kochanek C.S., Schechter P.L., 2004, Measuring and Modelling
the Universe, Carnegie Obs Centennial Symposium, ed. W. Freedman, CUP,
p. 117

\noindent Krist J., 1993. Astronomical Data Analysis Software and
Systems II, A.S.P. Conference Series, Vol. 52, 1993, R. J. Hanisch, R.
J. V. Brissenden, and Jeannette Barnes, eds., p. 536.

\noindent Marshall P.J., Hogg D.W., Moustakas L.A., Fassnacht C.D.,
Bradac M., Schrabback T., Blandford R.D., 2008, astro-ph/0805.1469

\noindent Matsumoto A., Futamase T., 2008, MNRAS 384, 843

\noindent Miralda-Escude J., Lehar J., 1992, MNRAS 259, 31P

\noindent Mobasher B., et al., 2007, ApJS 172, 117

\noindent Morgan C.W., Eyler M.E., Kochanek C.S., Morgan N.D., Falco
E.E., Vuissoz C., Courbin F., Meylan G., 2008,  ApJ 676, 80

\noindent Ofek E., Maoz D., 2003, ApJ 495, 101

\noindent Ofek E., Rix H.-W., Maoz D., 2003, MNRAS 343, 639

\noindent Peng C.Y., Ho L.C., Impey C.D., Rix H., 2002, AJ 124, 266.

\noindent Refsdal S. 1964,  MNRAS, 128, 307. 

\noindent Scoville N., et al., 2007, ApJS, 172, 1.

\noindent Seidel G., Bartelmann M. 2008,  AAS, 21116, 011. 

\noindent Taniguchi Y., et al., 2007, ApJS, 172, 9

\noindent Walsh D., Carswell R.F., Weymann R.J. 1979,  Nature, 279, 381. 

\end{document}